\documentclass[journal]{IEEEtran}
\usepackage{amsmath}
\usepackage{multicol}
\usepackage{amssymb}
\usepackage{graphicx}
\usepackage[utf8]{inputenc}
\graphicspath{{./Figures/}} 
\DeclareGraphicsExtensions{.pdf,.jpg,.png,.jpeg}
\usepackage[dvipsnames]{xcolor}
\usepackage{cite}

\title{{
~\vspace{-40pt}
\normalsize preprint April 2021;  to appear open access in IEEE Transactions on Power Systems} \\
\vspace{40pt}
Extracting resilience metrics from distribution utility data using  outage and restore process statistics
}
\author{Nichelle'Le K. Carrington,\,\IEEEmembership{Student Member,\,IEEE,} Ian Dobson,\,\IEEEmembership{Fellow,\,IEEE,}  Zhaoyu Wang,\,\IEEEmembership{Senior Member,\,IEEE}%
\thanks{
The authors are with ECpE Dept., Iowa State University, Ames IA 50011 USA; emails nkcarring,dobson,wzy@iastate.edu. We gratefully acknowledge support in part from NSF grants 1609080, 1735354, and 1549883.

\copyright 2021 Nichelle'Le Carrington and Ian Dobson and Zhaoyu Wang \newline Creative Commons Attribution License CC BY 4.0.
}%
}

\begin{document}

\maketitle

\begin{abstract}
Resilience curves  track 
the accumulation and restoration of outages during an event on an electric distribution grid.
We show that a resilience curve generated from utility data can always be decomposed into an outage process and a restore process and that these processes generally overlap in time.
We use many events in real utility data to characterize  the statistics of these processes, and derive formulas based on these statistics 
for resilience metrics such as restore duration, customer hours not served, and outage and restore rates.
The formulas express the mean value of these metrics as a function of the number of outages in the event.
We also give a formula for the variability of restore duration, which allows us
to predict a maximum restore duration  with 95\% confidence.
Overall, we give a simple and general way to decompose  resilience curves into outage and restore processes and 
then show how to use these processes to extract resilience metrics from standard distribution system data.
  \end{abstract}
  

\section{Introduction}

Under normal conditions,  component outages in electric power distribution systems occur at a low rate and are 
restored as they occur. However, when there is severe weather or other extreme stresses, the component outages occur at a high rate, and the outaged components accumulate until they are gradually restored.
This paper studies these resilience events in which outaged components accumulate.
We process 5 years of outage data recorded by a distribution utility to obtain many such real events for statistical analysis. 
Each event has a resilience curve that tracks the cumulative number of components or customers outaged  as they accumulate and then are restored. 

It is customary 
to divide resilience curves into successive, non-overlapping phases.
For example, Nan \cite{NanRESS17}  describes a disruptive outaging phase followed by a recovery phase,
while Panteli \cite{PanteliProcIEEE17} and Ouyang \cite{OuyangSS12} describe a resilience trapezoid with the three phases of progressive disturbances, then a degraded or assessment phase, then recovery in a resilience triangle.
Instead of a resilience triangle with a straight line hypotenuse, Reed \cite{ReedSYS09} describes restoration of power line outages with an exponential curve.
Similarly,
Yodo \cite{YodoASMEJMD16} describes resilience curves in terms of successive unreliability, disrupted, and recovery phases with triangles, trapezoids, and other curves.
Carrington \cite{CarringtonPESGM20} uses the nadir of resilience curves from utility data  to separate an outage phase from a recovery  phase.
Many other papers  have similar accounts of resilience phases.

These distinct phases of resilience are conceptually compelling.
Moreover, the dimensions, slopes, and areas of the resilience triangles and trapezoids 
define standard resilience metrics of event duration, average rates of outage or recovery, and overall impact.

However, working with our utility data suggests a different point of view in which outage and restore processes  routinely overlap in 
time. Indeed, the average fraction of event duration for which outage and restore processes overlap\footnote{that is, the average of $(o_n-r_1)/(r_n-o_1)$ in the notation of section~\ref{outagerestore}} is 
0.61 for events with 10 to 20 outages, 
0.89 for events with 100 to 200 outages, 
and
0.95 for events with 1000 to 2000 outages.
That is, for practical processing of real distribution system outage data we decompose the resilience curve not into successive phases but into outage and restore processes that occur together for much of the time. 
Moreover, we will show that standard resilience metrics can still be obtained with this approach.
Our point of view is general: outage and restore processes can always be combined into a resilience curve, and 
any resilience curve
can always be decomposed into an outage process and a restore process.

We start by extracting from the utility data many resilience events and their  resilience curves and the outage and restore processes. These track the number of components out during each event.
Then we obtain the statistics of the outage and restore processes and derive the  resilience metrics.
The outage process is statistically characterized by the times between successive outages or the outage rate.
The restore process starts after a delay and is statistically characterized by the times between successive restores or the restore rate. 
The duration of the restore process and of the entire event are then easily obtained standard metrics.

There is also a conventional customer resilience curve tracking the  number of customers out during the event that we also obtain from the utility data.
This customer resilience curve can also be decomposed into a
 customer outage process and a customer restore process. 
We can measure the impact of an event by the customer hours lost, which is  the area under the customer resilience curve
and a well-known resilience metric \cite{NanRESS17,PanteliProcIEEE17,YodoASMEJMD16}.
We compute the mean customer hours lost from the statistics of the customer outage and restore processes.

Previous pioneering work on queueing models of reliability and resilience has used outage and restore processes. Zapata \cite{ZapataTandDCE08} models distribution system reliability with outages as a point process arriving at a queue that is serviced by a repair process with multiple crews to produce an output that is a restore process. 
Wei and Ji \cite{WeiAM16} analyze distribution system resilience to particular severe hurricanes with an outage process  arriving at a queue with a repair process to produce a restore process. 
In \cite{WeiAM16}, these processes vary in both time and space as the hurricane progresses.
Both \cite{ZapataTandDCE08} and \cite{WeiAM16} statistically model the outage process and the repair process of components,  and then calculate the restore process.
With our focus on studying the overall system resilience with real data, we can model the restore process directly from the data, and avoid the  complexities of explicitly modeling the repair of components and assuming an order in which they are repaired.

Our methods require a sufficient number of events for good statistics, so that this paper addresses the more common, less extreme events.
Therefore there is little overlap of this paper with \cite{WeiAM16}, which  addresses individual instances of the most extreme events (direct hits by a hurricane), for which there are few events for a given utility.

There are several methods of estimating the number of outages in an anticipated storm \cite{LiuPS07,LiuJIS05,KankanalaPS14,ZhuEPSR07,ZhouPS06,AlvehagPD11,LiIBM10},
 including practical utility application in \cite{LiIBM10}.
Since some of our statistics show a dependence on the number of outages,
this capability to predict the number of outages will be useful in applying the results in this paper to anticipated storms.

The utility data also yield estimates of the variability of the outage and restore processes, enabling  estimates of the variability of the restore duration.
Then, given the estimate of the number of outages, we can use our 
restore time statistics to 
 predict upper bounds of the restore duration of an anticipated storm, such as its 95th percentile. 
The upper bound is intended to help the utility predict when the 
restore process will be completed with more confidence.

With a similar overall aim,
 previous work estimates individual component restoration times  from utility data in different ways.
For example, Jaech \cite{JaechPS19} predicts a gamma distribution of  individual component outage restoration times and customer hours lost with a  neural network that processes utility  records and wind speed, outage time and date, and hence obtains upper bounds of individual component restoration times.
Chow \cite{ChowPD96} analyzes the contributions of timing, faults, protection, outage types and weather to individual component restoration times.
Liu \cite{LiuRESS08} fits generalized additive accelerated failure time models to hurricane and ice storm utility data. The individual outages were then combined to give system restoration curves at the county level.
Liu \cite{LiuRESS08} and Maliszewski \cite{MaliszewskiAG12}  model the spatial variation of average outage times.

We summarize the innovations in the paper:

(1) We extract from 5 years of utility data many events in which multiple outages accumulate before being restored and perform a novel statistical analysis of these resilience events.

(2) We separate the events into overlapping outage and restore processes for both  components and  customers, and obtain 
 statistics of these processes. The restore process is directly characterized from the data without modeling repair of individual  components.

(3) We derive formulas for the mean and standard deviations of restore and event durations
from the statistics of the outage and restore processes, and illustrate their application to estimating an upper bound for the restore duration. We derive a formula for the mean customer hours lost.

\vspace{3pt}
\noindent
These innovations allow us to extract and compute standard metrics for resilience events from practical utility data.

\section{Outage and restore processes}
\label{outagerestore}

\begin{figure}[htb]
	\centering
	\includegraphics[width=\columnwidth]{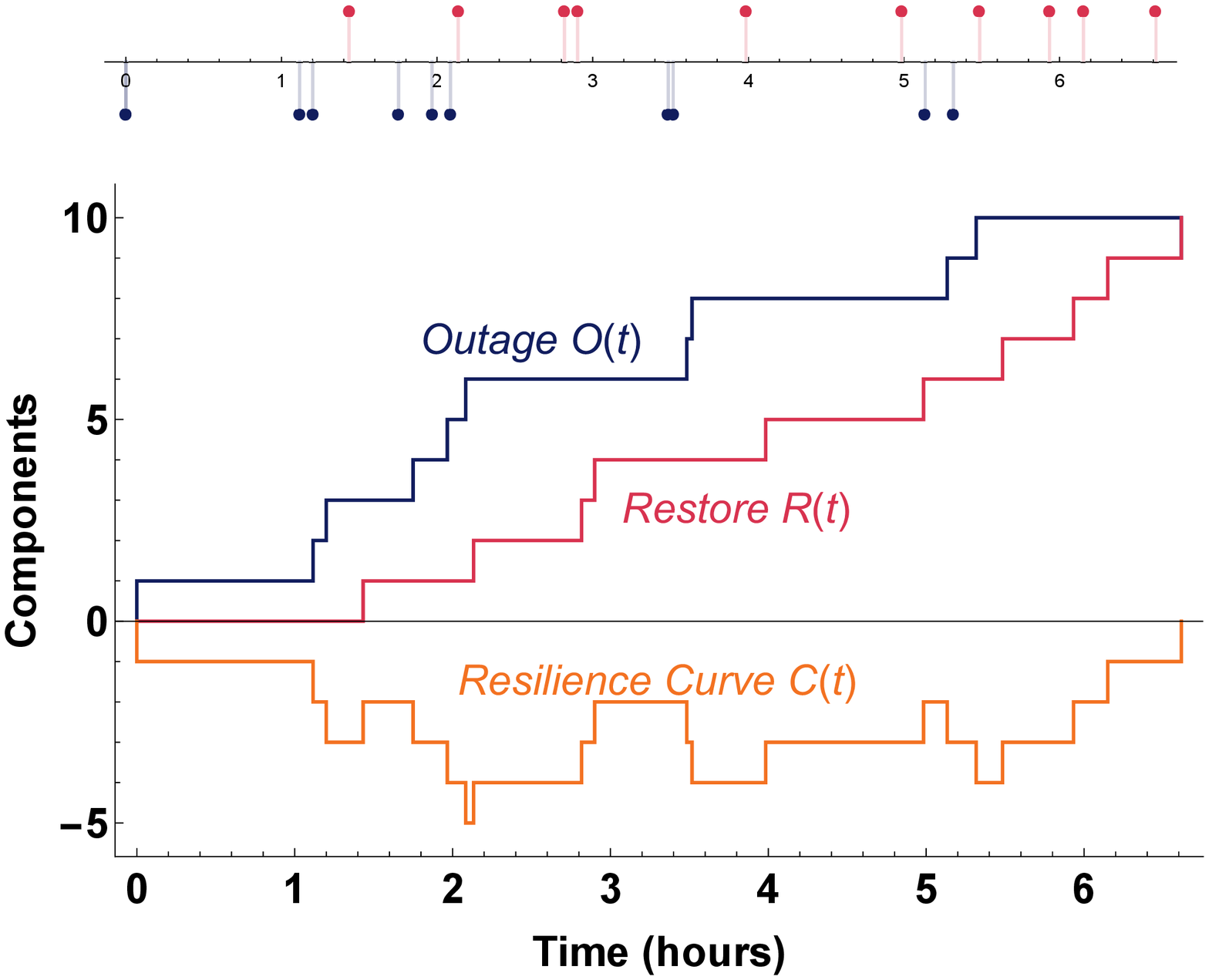}
	\caption{A component resilience curve and its associated outage and restore processes.}
	\label{ComponentProcesses}
\vspace{10pt}
	\centering
	\includegraphics[width=\columnwidth]{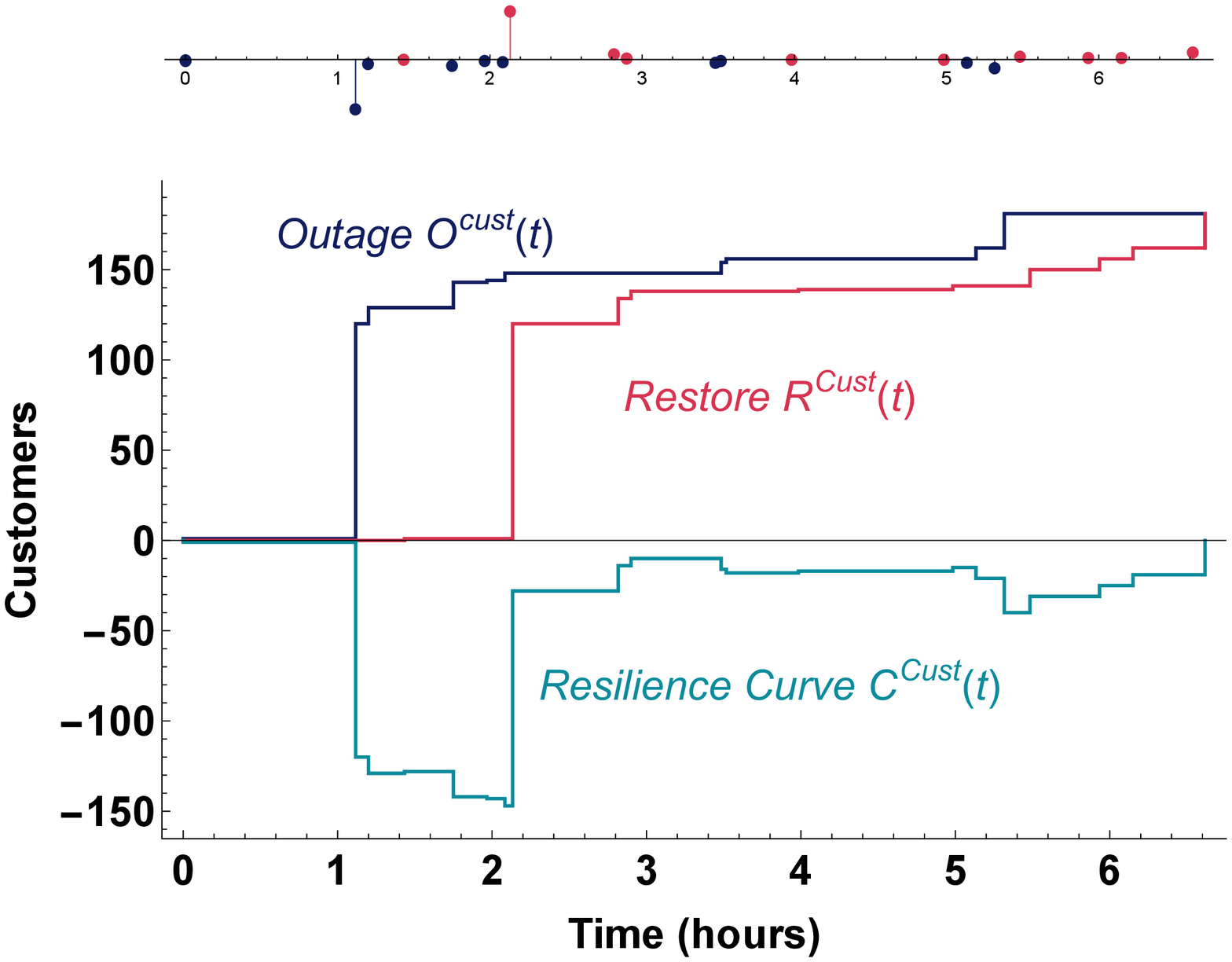}
	\caption{A customer resilience curve and its associated customer outage and customer restore processes for the same event  as Figure \ref{ComponentProcesses}.}
	\label{CustomerProcesses}
\end{figure}

The resilience events of interest occur  when outaged components accumulate before being restored. Each event has a conventional resilience curve $C(t)$ for component outages.  $C(t)$ is the negative of the 
cumulative number of component outages as a function of time $t$.  For example, the orange curve at the bottom portion of Figure \ref{ComponentProcesses} shows $C(t)$ for an event with 10 component outages. 
We now explain the outage and restore processes  and how they relate to the resilience curve.

\begin{figure*}[htb]
	\centering
	\includegraphics[width=\textwidth]{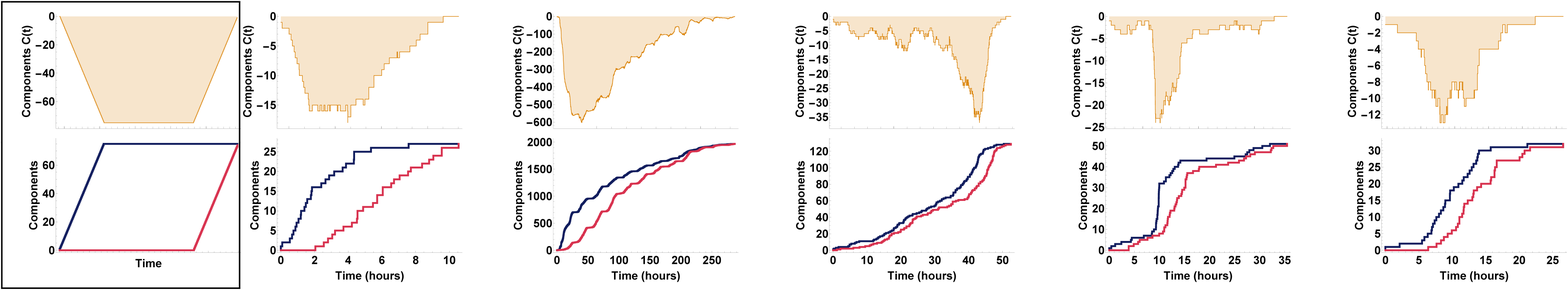}
		\vspace{9pt}
		
	\includegraphics[width=\textwidth]{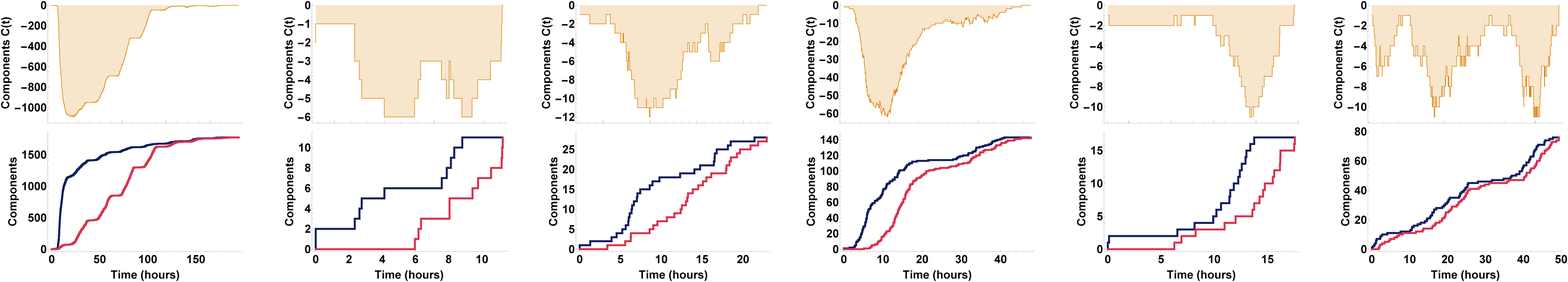}
	
	\caption{Component resilience curves (upper rows with one shaded curve) and their corresponding decompositions into outage and restore processes (lower rows with blue and red curves). The first example is an idealized case with trapezoidal resilience curve and all the rest are examples from utility data.}
	\label{decomposeexamplefig}
	\vspace{-10pt}
\end{figure*}
\subsection{Examples of component outage and restore processes}
\label{examples}

We start with no components outaged. 
Then 10 components outage at times $o_1\le o_2\le ...\le o_{10}$ as shown by the tick marks below the top time line of Figure \ref{ComponentProcesses}. 
The restore times are $r_1\le r_2\le...\le r_{10}$ as shown by the tick marks above the top time line of Figure \ref{ComponentProcesses}.
The restore times are numbered in the time order that they occur.
The cumulative number of outages $O(t)$ at time $t$ and the 
cumulative number of restores $R(t)$ at time $t$ are
defined by counting $1$ for each outage or restore before time $t$:
\begin{align}
O(t)&=\sum_{\displaystyle k \mbox{ with } o_k\le t}\hspace{-4mm}1
\label{cumulativeoutages}\\
R(t)&=\sum_{\displaystyle k \mbox{ with } r_k\le t}\hspace{-4mm}1
\label{cumulativerestores}
\end{align}
Figure \ref{ComponentProcesses} shows the cumulative number of outages $O(t)$ and the cumulative number of restores $R(t)$.
In this case, $O(t)$ and $R(t)$ increase from zero to the total number of outages 10.

The cumulative number of component outages at time $t$ is $O(t)-R(t)$.
The component resilience curve $C(t)$ is defined as the negative of the cumulative number of component outages at time $t$ so that
\begin{align}
C(t)=R(t)-O(t)
\label{componentresiliencecurve}
\end{align}
Figure \ref{ComponentProcesses} shows how the resilience curve $C(t)$ can be decomposed into the restore process minus the outage process.

It is clear from (\ref{componentresiliencecurve}) that any outage and restore processes $O(t)$ and $R(t)$ define a resilience curve $C(t)$.
Moreover,
any resilience curve $C(t)$ can be uniquely decomposed as (\ref{componentresiliencecurve}) into outage and restoration processes $O(t)$ and $R(t)$ that increase from zero to the number of outaged components $n$. 
In mathematics this decomposition is known as the Jordan decomposition \cite{EOM} of functions of bounded variation\footnote{The total variation of $C(t)$ is $2n$, which is bounded. In our case the Jordan decomposition (\ref{componentresiliencecurve}) is minimal and unique since we require that $O(t)=R(t)=0$ for $t<o_1$ and $O(t)+R(t)=2n$ for $t>r_n$ \cite[defn.~2.4(2), thm.~2.5(2)] {JafarikhahLMCS14}, \cite[sec. 9-4]{Taylor}. 
Since $C(t)$ is minus the cumulative number of outages,
if there are simultaneous restores and outages, for example,  $m_r$ restores and $m_o$ outages all occurring precisely at time $t$, then only their difference $m_r-m_o$ contributes to $C(t)$, and we assume that only $|m_r-m_o|$ contributes to the total number $n$ of outages or restores. Note that 
$O(t),R(t),C(t)$ are right continuous.}.

There is noticeable  variety in the forms of the component resilience curves in our utility data.  The examples (except for the first example) in Figure~\ref{decomposeexamplefig}
 show these curves and their decompositions into outage and restore processes. For comparison, the first example in 
 Figure~\ref{decomposeexamplefig} shows a conventional, idealized case of a trapezoidal resilience curve.

Considering the outage  and  restore processes separately is useful because 
they correspond to different aspects of system resilience: the outage process  results from
individual component strengths under bad weather stress and the restore process  results from the control room, restoration plans, and the number and performance of restoration crews.

\subsection{Extracting events from utility data}  
The historical outage data were recorded by one distribution utility from 2011 to 2016 over a territory including rural areas and small cities.
32\,291 outages were recorded during this time.   
The start and end time of the outages were recorded by fuse cards based on the loss of power at that location.
The number of customers out for each outage was also recorded. 

We use our method from \cite{CarringtonPESGM20}  to define events.
The start of an event is defined by an initial outage that occurs when all components are functional, and the end of the same event is defined by the first subsequent time when all the components are restored. That is, the event starts when the cumulative number of failures $C(t)$ first changes from zero and ends when $C(t)$ returns to zero.

In particular, we sort the combined component outage
and restore times by their order of occurrence and then calculate the cumulative number of outages 
$C(t)$ at all the outage and restore times. 
Each restore at time $r_n$ for which $C(r_n)=0$  is the end of an event and the immediately following outage is the start of the next event.
Note that if the event has $n$ outages, then it must have $n$ restores to allow the cumulative number of outage $C(t)$ to return to zero at time $t=r_n$. 
 
Applying this event processing to the historical data yields 2618 events.  
The component and customer resilience curves for each event are decomposed into  outage and restore processes as explained in more detail in the next subsection.

\subsection{Component outage and restore processes}
\label{processes}

This subsection explains the 
outage and restore processes in more detail and shows how their statistics are extracted from the events in the utility data.

\begin{figure}[t]
	\centering
	\includegraphics[width=\columnwidth]{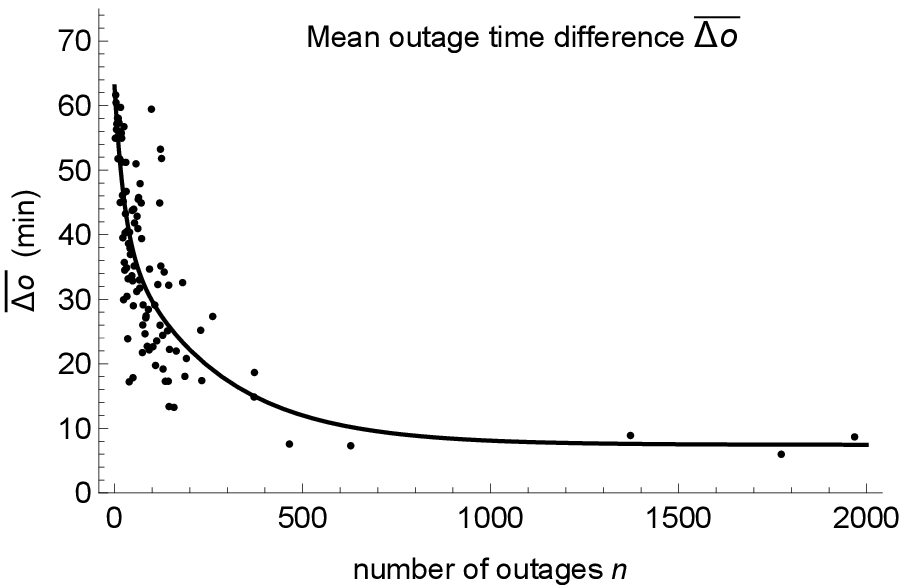}
\vspace{5mm}
	\includegraphics[width=\columnwidth]{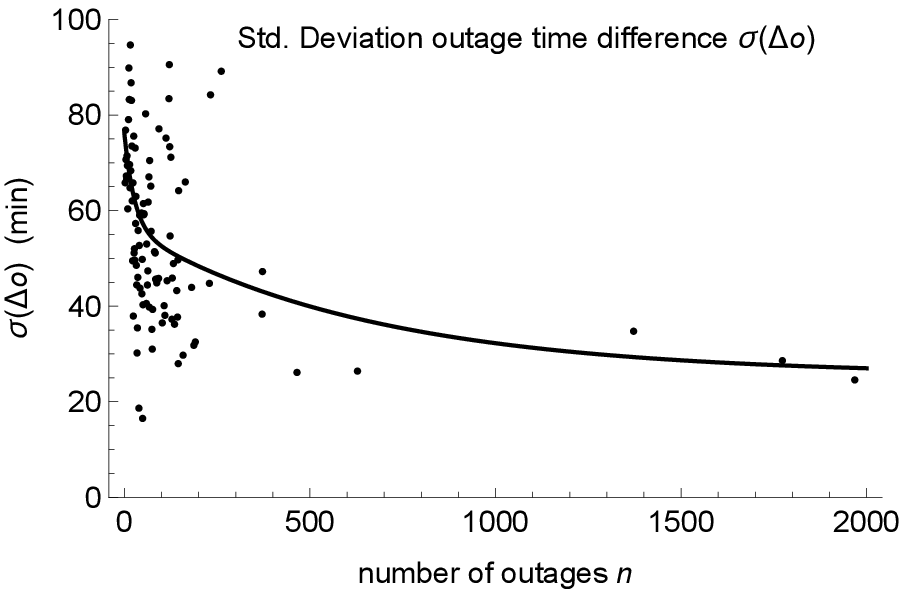}
	\caption{Mean and standard deviation of outage time difference empirical data (dots) and fitted curve as a function of number of outages $n$.}
	\label{outagefit}
	\vspace{-5mm}
\end{figure}

Suppose that $o_1\le o_2\le...\le o_n$ are the component outage times in an event in order of occurrence and that $\Delta o_k=o_{k+1}-o_k$, $k=1,...,n-1$ are the times between successive component outages.
The outage time differences $\Delta o_k$, $k=1,...,n-1$ can be regarded as independent samples from a 
probability distribution of outage time differences  $\Delta o$.
We want to find the mean and standard deviation of $\Delta o$ as a function of $n$.

To do this, we combine the outage time differences $\Delta o_k$ for all the events with $n$ outages and calculate their mean and standard deviation. 
Then we fit the empirical mean and the standard deviation as  functions of $n$ with a linear combination of a constant and 2 exponential functions to smooth and interpolate the data as shown in Figure~\ref{outagefit}.
 The constant plus exponentials  fitting function is chosen as a simple form to match the way the average data decrease towards a positive value.
The functional fits are 
\begin{align}
\overline{\Delta o}&=7.45 +23.3 e^{-0.0388 n}+32.2 e^{-0.00391 n}\quad{\rm min}\\
\sigma(\Delta o)&=25.6 +19.5 e^{-0.0375 n}+30.9 e^{-0.00153 n}\quad{\rm min}
\end{align}

\begin{figure}[t]
	\centering
	\includegraphics[width=\columnwidth]{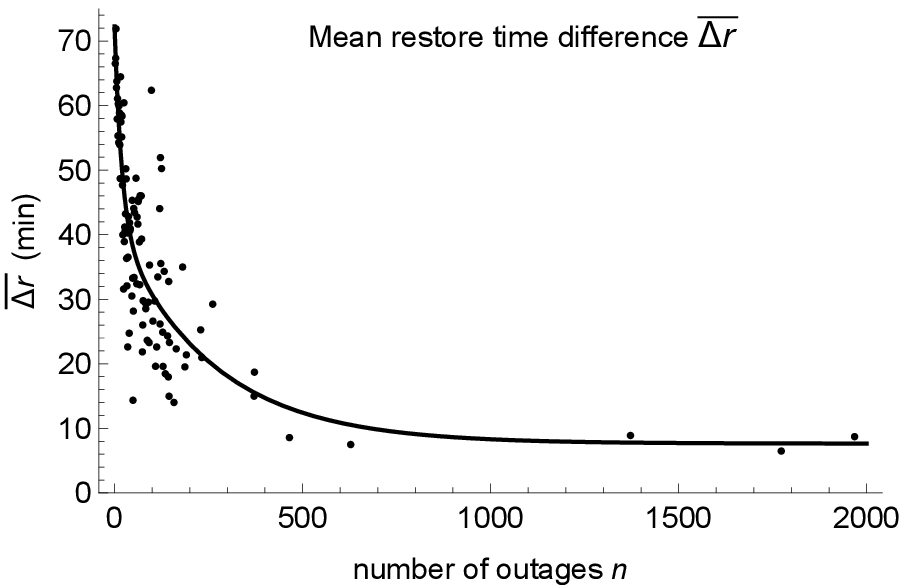}
\vspace{5mm}
	\includegraphics[width=\columnwidth]{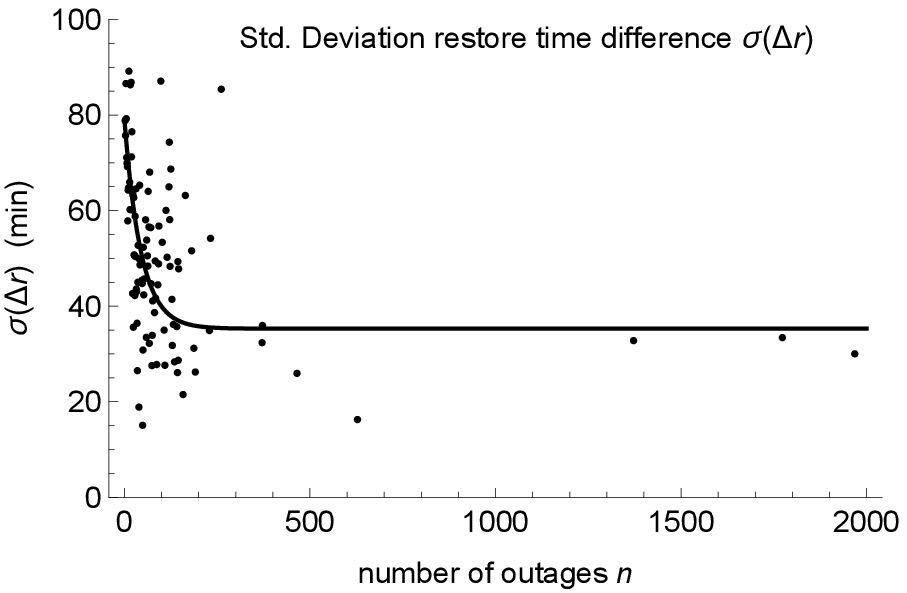}
	\caption{Mean and standard deviation of restore time difference empirical data (dots) and fitted curve as a function of number of outages $n$.}
	\label{restorefit}
	\vspace{-5mm}
\end{figure}
Suppose that $r_1\le r_2\le...\le r_n$ are the component restore times in order of occurrence.
Note that the component outaged in the $k$th outage can be different from the component restored in the $k$th restore.
In effect, we disregard {\sl which} component is restored and only track that {\sl some} component is restored.
(We call $r_1, r_2,...,r_n$ component restore times to minimize any confusion with the restoration or repair times of particular components.)

Let $\Delta r_k=r_{k+1}-r_k$, $k=1,...,n-1$ be the times between successive component restores.
The restore time differences $\Delta r_k$, $k=1,...,n-1$ can be regarded as independent samples from a 
probability distribution of restore time differences  $\Delta r$.
We extract the statistics of  $\Delta r$ from the utility data as a function of the number of outages $n$ similarly as $\Delta o$.
Figure~\ref{restorefit} plots the mean restore time difference $\overline{\Delta r}$ and the standard deviation of the restore time difference $\sigma(\Delta r)$ and the functions fitted.
The functional fits are 
\begin{align}
\overline{\Delta r}&=7.64 +30.8 e^{-0.0514 n}+33.8 e^{-0.00391 n} \quad{\rm min}\label{meandelatr}\\
\sigma(\Delta r)&=35.3 +43.7 e^{-0.0224 n}\quad{\rm min}
\label{SDdelatr}
\end{align}

Let $\Delta r_0=r_1-o_1$ be the delay in the start of the restoration process relative to the start of the event at time $o_1$. 
One factor contributing to $\Delta{r_0}$ is utility inspection crews and clean-up crews working to ensure the safety of the area and assess the damage needing repair.
There is no clear trend in the variation of $\Delta r_0$ with $n$, so we combine the data for $\Delta r_0$ for all events with 2 or more outages and 
compute 
\begin{align}
\overline{\Delta r_0}=132\ {\rm min}
\qquad\mbox{and}\qquad
\sigma(\Delta r_0)=92.4\ {\rm min}
\label{firstdelay}
\end{align}

 The observed events become sparse as the number of outages $n$ increases as shown in Table \ref{tablemeansamples}.
    However, the number of samples of times between restores $\Delta r$  and the times between outages $\Delta o$ available in the data for number of outages $n$ is proportional to both $n-1$  and the observed number of events with $n$ outages.
    Table \ref{tablemeansamples} shows that
    there are more samples of $\Delta r$ and $\Delta o$ available in the data both for the small outages with many events and the largest outages with only one event. This improves the estimation of $\Delta r$ and $\Delta o$ for the smallest and largest events.
   
     \begin{table}[ht]
       \centering
       \caption{Numbers of samples of time differences and  events}
       \begin{tabular}{@{}cccc@{}}
          \text{range of $n$} & \text{mean \# samples of} & \text{mean \# events} & \text{total \# events}\\
          \text{($n$ = \# outages)} & \text{$\Delta r\,\&\,\Delta o$  for each $n$} & \text{ for each $n$} & \text{ in range of $n$}\\
          \hline
 2--10 & 738 & 238.8 &2149\\
 11--20 & 367 & 27.7 &277\\
 21--30 & 173 & 7.2 &72\\
 31--40 & 112 & 3.3 &30\\
 41--50 & 106 & 2.4 &17\\
 51--60 & 127 & 2.3 &14\\
 61--70 & 74 & 1.2 &7\\
 71--80 & 88 & 1.2 &6\\
 81--90 & 118 & 1.4 &7\\
 91--100 & 92 & 1.0 &2\\
 101--120 & 110 & 1.0 &6\\
 121--200 & 158 & 1.1 &20\\
 201--1000 & 364 & 1.0 &7\\
 1001--2000 & 1703 & 1.0 &3\\
  \hline
       \end{tabular}
       \label{tablemeansamples}
   \end{table}

\subsection{Customer outage and restore processes}
\label{customerprocess}
The utility data record the number of customers outaged for each outage, allowing us to similarly analyze  resilience curves tracking the number of customers out  to obtain outage and restore processes for the number of customers.
Generalizing (\ref{cumulativeoutages}) and (\ref{cumulativerestores}),
the cumulative numbers of customers out $O^{\rm cust}(t)$ and customers restored $R^{\rm cust}(t)$ at time $t$ are
\begin{align}
O^{\rm cust}(t)&=\sum_{\displaystyle k \mbox{ with } o_k\le t}\hspace{-4mm}c_k^{\rm out}
\label{cumulativeoutagescust}\\
R^{\rm cust}(t)&=\sum_{\displaystyle k \mbox{ with } r_k\le t}\hspace{-4mm}c_k^{\rm res}
\label{cumulativerestorescust}
\end{align}
The customer resilience curve $C^{\rm cust}(t)$ is now obtained similarly to the component resilience curve (\ref{componentresiliencecurve}) as
\begin{align}
C^{\rm cust}(t)=R^{\rm cust}(t)-O^{\rm cust}(t).
\label{customerresiliencecurve}
\end{align}
Figure \ref{CustomerProcesses} shows an example of  customer processes $O^{\rm cust}(t)$ and $R^{\rm cust}(t)$ and
 the resilience curve $C^{\rm cust}(t)$.
The processes $O^{\rm cust}(t)$ and $R^{\rm cust}(t)$ increase from zero to the total number of customers out, which is 181 in this example.

When an outage that disconnected customers is restored, the same number of customers are restored. However, outages are not necessarily restored in the order that the outages occurred. 
Therefore the numbers of customers restored $c_1^{\rm res}, c_2^{\rm res},..., c_n^{\rm res}$ are a  permutation of the numbers of customers out $c_1^{\rm out}, c_2^{\rm out},..., c_n^{\rm out}$.

The numbers of customers out $c_1^{\rm out}, c_2^{\rm out},..., c_n
^{\rm out}$ can be regarded as $n$ independent samples from a distribution $c$ of number of customers out. The customers restored  $c_1^{\rm res}, c_2^{\rm res},..., c_n
^{\rm res}$ can also be regarded as $n$ independent samples from  $c$.
We combine the data for customers out for all events\footnote{2\% of the customer data are blank entries that we replaced with 0.}
and 
compute the mean and standard deviation of the number of customers:
\begin{align}
\overline{c}=54.0
\qquad\mbox{and}\qquad
\sigma(c)=180.
\end{align}
The customers out for each outage are determined by the location of the outage in the network, and the distribution of the number of customers out is determined by the overall network design and its vulnerabilities.

\section{Resilience metrics} 

We express event durations in terms of the time differences of the restore  process and then derive formulas for the mean and standard deviations of the restore duration and the event duration. We also combine the time differences with the customers outage statistics to derive a  formula for the mean customer hours lost. The average outage and restore rates are obtained.

The restore process starts at time $r_1$ and ends at time $r_n$, so the 
restore duration is
    \begin{align}
    D_R&=r_n-r_1
=(r_2-r_1)+(r_3-r_2)+...+(r_n-r_{n-1})\notag\\
    &= \Delta r_1+ \Delta r_2+...+ \Delta r_{n-1}
\label{DR}
\end{align}
The mean restore duration is then
\begin{align}
\overline{D_R}&=
  (n-1)\overline{\Delta r}
    \label{meanDR}
\end{align}
Assuming that   $\Delta r_1$, $\Delta r_2$, ..., $\Delta r_{n-1}$ are independent, we obtain
$\sigma^2(D_R) = (n-1)\,\sigma^2(\Delta r)$ and
 \begin{align}
       \sigma(D_R) &= \sqrt{n-1}\,\sigma(\Delta r)
     \label{SDR}
    \end{align}
    In (\ref{DR}), the restore duration $D_R$ is measured until the last restore of the event. If it is preferred to measure the restore duration until, say, 95\% of the outages are restored,
then this can easily be done by replacing $n-1$ in (\ref{meanDR}) and (\ref{SDR}) by $\lceil 0.95\, n -1 \rceil$, where the ceiling function $\lceil \cdot \rceil$ rounds up to the nearest integer.

Each event starts at the first outage time $o_1$ and ends at the last restore time $r_n$. Then the event duration is 
\begin{align}
D_E&=r_n-o_1
=(r_1-o_1)+(r_n-r_1)
  =\Delta r_0+D_R
    \label{D}
\end{align}
Using (\ref{meanDR}), the mean event duration is 
\begin{align}
\overline{D_E}&=
    \overline {\Delta r_0}+ (n-1)\overline{\Delta r}
    \label{meanD}
\end{align}
and, since 
$\Delta r_0$ and $D_R$ are independent, we use (\ref{SDR}) to obtain
$ \sigma^2(D_E) = \sigma^2(\Delta r_0) +(n-1)\sigma^2(\Delta r)$ and
 \begin{align}
      \sigma(D_E) &= \sqrt{\sigma^2(\Delta r_0) +(n-1)\sigma^2(\Delta r)}
     \label{SDD}
    \end{align}

\begin{figure}[htb]
	\centering
	\includegraphics[width=\columnwidth]{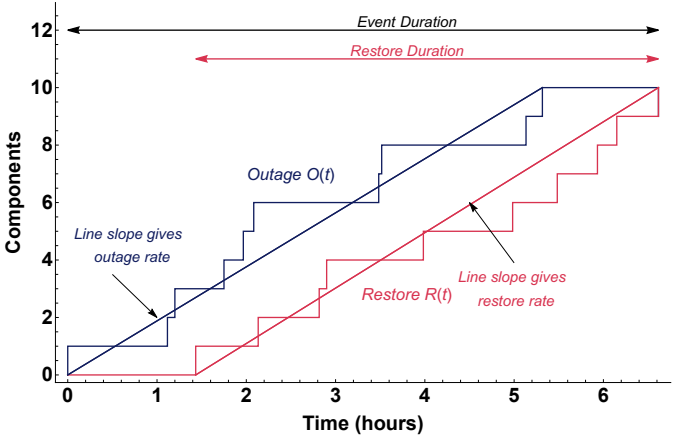}
	\caption{Resilience metrics (durations and rates) for the component outage and restore processes.}
	\label{ComponentDiagram}
\end{figure}

The restore and outage rates during events\footnote{\looseness=-1
The  outage rate measured over a year is much lower than the  outage rate during events because it accounts for the time between events.} are
\begin{align}
\lambda_R &=(\Delta r)^{-1}
\label{lambdaR}\\
\lambda_O &=(\overline{\Delta o})^{-1}
\label{lambdaO}
\end{align}
In (\ref{lambdaR}) we obtain the restore rate $\lambda_R $ from $\overline{\Delta r}$, which is a quantity averaged over events. 
This restore rate $\lambda_R $  should be distinguished from the instantaneous restore rate $\lambda_R^{\rm inst}(t)$, which has been observed in \cite{WeiAM16} to vary with time for the largest events. 
 The restore rate $\lambda_R $ averaged over events can usefully apply even as the instantaneous restore rate varies\footnote{Consider the idealized case of a nonhomogeneous Poisson  recovery process. 
 Suppose there are 3 restoring processes on the time intervals $D^{(1)}$, $D^{(2)}$, $D^{(3)}$, each of duration $T$ and with $n$ restores.
 Then the expected value of $\int_{t\in D^{(i)}}\lambda_R^{\rm inst}(t)dt$ is $n-1$ and the duration $T= \sum_{k=1}^{n-1} \Delta r_k^{(i)}=(n-1) \overline{\Delta r}^{(i)}$ for $i=1,2,3$.
 Then we estimate 
 $\lambda_R =\frac{1}{3T}\sum_{i=1}^3\int_{t\in D^{(i)}}\lambda_R^{\rm inst}(t)dt$ as $\frac{3(n-1)}{(n-1) (
 \overline{\Delta r}^{(1)}+\overline{\Delta r}^{(2)}+\overline{\Delta r}^{(3)})
 }= (\overline{\Delta r})^{-1}$.
 }.

The restore process depends largely on the restoration capability available to the utility. 
The restore process metrics are  $\overline{\Delta r_0}$ and $\overline{D_R}$ or $\lambda_R$.
Increasing the number of utility crews or their effectiveness would decrease $\overline{\Delta r_0}$ and $\overline{D_R}$, and increase $\lambda_R$.

The outage process depends on a combination of the weather impact and the condition and strength
of the grid components.
The number of outages $n$ will vary with the weather and the condition of the grid components, increasing if the weather is more extreme or more prolonged, or if the grid components are weaker.

\begin{figure}[htb]
	\centering
	\includegraphics[width=\columnwidth]{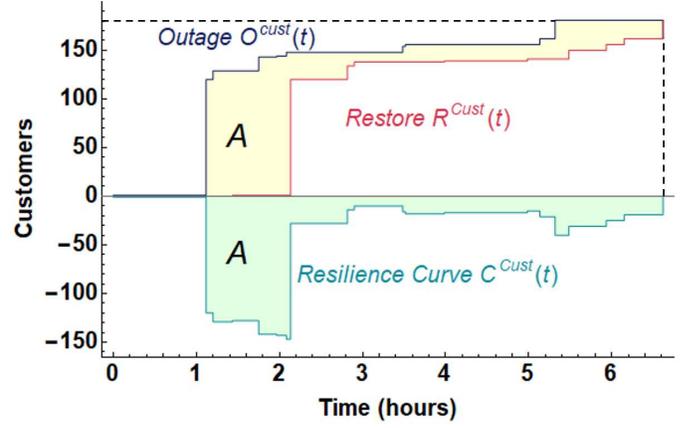}
	\caption{ Area $A$ under resilience curve is the customer hours metric and is equal to the area $A$ between the outage and restore processes.}
	\label{CustomerFilledArea}
\end{figure}

We use  the customer hours lost  $A$ to quantify
the customer impact of an event.
$A$ is the area under the customer resilience curve:
\begin{align}
A=-\int_{
\displaystyle o_1}^{\displaystyle r_n}C^{\rm cust}(t)dt
\label{customerimpact}
\end{align}
The minus sign in (\ref{customerimpact}) makes $A$ a positive area.
Using (\ref{customerresiliencecurve}),
 $A$ is also the area between the customer outage and restore curves: 
\begin{align}
A=\int_{
\displaystyle o_1}^{\displaystyle r_n}\big[O^{\rm cust}(t)-R^{\rm cust}(t)\big]dt
\label{A}
\end{align}
The two interpretations of area $A$ are illustrated in Figure \ref{CustomerFilledArea}.

Consider the rectangle indicated by the dashed lines and the axes in Figure  \ref{CustomerFilledArea}.
Let $A_R$ be the area in the rectangle above the restore curve and let $A_O$ be the area in the rectangle above the outage curve.
Then, since the customer hours lost $A$ is the area between the outage curve and the restore curve,
\begin{align}
A&=A_R-A_O,\\
\intertext{where} 
A_R&=\bigg(\sum_{j=1}^n c_{j}^{\rm res}\bigg)\Delta r_{0}  +\sum_{i=2}^{n}\sum_{j=i}^n c_{j}^{\rm res}\Delta r_{i-1}
\label{ARR}\\
A_O&=\sum_{k=2}^{n}\sum_{\ell=k}^n c_{\ell}^{\rm out}\Delta o_{k -1}
\label{AOO}
\end{align}
Since the restore and outage curves are piecewise constant, 
the expressions for the areas $A_R$ and $A_O$ in
(\ref{ARR}) and (\ref{AOO}) were obtained  by summing the areas of the rectangles under the curves (recall the Riemann sum definition of an integral). 

\begin{figure}[htb]
	\centering
	\includegraphics[width=\columnwidth]{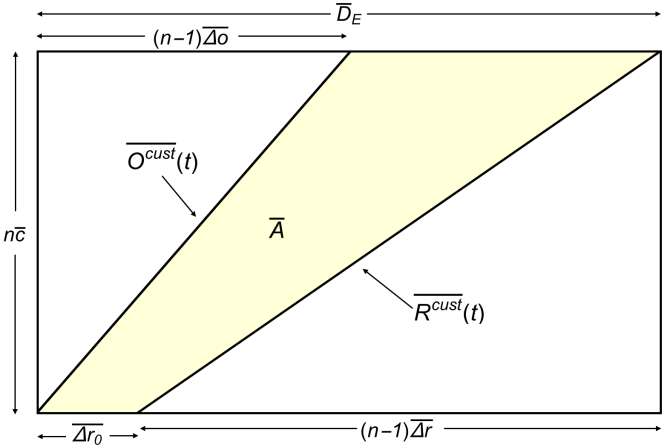}
	\caption{Averaged dimensions and customer outage and restoring processes shown to calculate the customer hours $\overline A$.}
	\label{AreaVariables}
\end{figure}

Since the customers out are independent of the times between restores or outages (so that expectations of products are products of expectations) and using $\sum_{i=2}^{n}\sum_{j=i}^n1=\textstyle{\frac{1}{2}}n(n-1)$, we take expectations of (\ref{ARR}) and (\ref{AOO}) to get 
\begin{align}
\overline{A_R}&=\bigg(\sum_{j=1}^n \overline{c}\bigg)\overline{\Delta r_0}  +\sum_{i=2}^{n}\sum_{j=i}^n \overline{c}\,\overline{\Delta r}\notag\\&=
n\overline{c}\,\overline{\Delta r_0}  +{\textstyle\frac{1}{2}}n(n-1) \overline{c}\,\overline{\Delta r}
\label{ARRmean}\\
\overline{A_O}&=\sum_{k=2}^{n}\sum_{\ell=k}^n \overline{c}\,\overline{\Delta o}={\textstyle\frac{1}{2}}n(n-1) \overline{c}\,\overline{\Delta o}
\label{AOOmean}
\end{align}
Hence, 
\begin{align}
\overline{A}&=\overline{A_R}-\overline{A_O}
=n\overline{c}\,\overline{\Delta r_0}+ {\textstyle\frac{1}{2}}n(n-1)\overline{c}(\overline{\Delta r}-\overline{\Delta o})
\label{Amean1}
\end{align}
An alternative expression for (\ref{Amean1}) can be obtained using  (\ref{meanD}):
\begin{align}
\overline{A}&=n\overline{c}\overline{D_E}- {\textstyle\frac{1}{2}}n(n-1)\overline{c}(\overline{\Delta r}+\overline{\Delta o})\label{Amean2}
\end{align}
The terms in  (\ref{ARRmean}) and (\ref{AOOmean}) can be understood by examining the corresponding areas in Figure \ref{AreaVariables}.
For example, the area $\overline{A_R}$ above the average restore curve  is the area of the rectangle with sides $\overline{\Delta r_0}$ and $n\overline{c}$  plus the area of the triangle with sides 
$(n-1)\overline{\Delta r}$ and $n\overline{c}$. 
And the area $\overline{A_O}$ above the average outage curve  is the area of the triangle with sides 
$(n-1)\overline{\Delta o}$ and $n\overline{c}$. 

It is useful to describe the outage and restore processes with separate parameters and separate metrics because they respond to different resilience investments. 
For example, a program of renewing or strengthening components would affect the outage process whereas an increased number of repair crews would affect the restoration process.
In more detail, (\ref{meanD}) shows the  effects on the average event duration of 
reducing the number of outages $n$ (by hardening the infrastructure) reducing $\overline{\Delta r_0}$ (by deploying more inspection crews) and reducing the average time between restores $\overline{\Delta r}$ (by deploying more repair crews). 
For a larger event, $n$ is larger and reducing $\overline{\Delta r}$  will have a larger effect because $\overline{\Delta r}$ is multiplied by $n-1$.
Formula (\ref{Amean1}) shows the corresponding effects on the customer hours $\overline A$. Hardening the upstream system or installing more reclosers can reduce the average customers disconnected per outage $\overline c$ and proportionally reduce $\overline A$. Reducing the number of outages $n$ or $\Delta r$ has an even greater effect for larger events because  the second term of (\ref{Amean1}) grows like $n^2$.

\section{Results}
  
This section gives numerical results illustrating the application of the formulas for the statistics of the metrics.

We can evaluate restore duration mean $\overline{D_R}$ and standard deviation $\sigma(D_R)$ for a given number of outages from (\ref{meandelatr}) and (\ref{SDdelatr}).
For example, if there are $n=10$ outages then 
 the restore duration has mean 527 min and  standard deviation  211 min.  
 If there are $n=100$ outages, then 
 the restore duration has mean 3038 min and  standard deviation  397 min.
 If these formulas are to be used for predicting restoration duration for an incoming storm, then the number of outages $n$ can be predicted by a number of methods as reviewed in the introduction.
 As well as estimating the mean,
 it is useful in applying the restore duration to compute its variability with its standard deviation.

  The event duration $D_E$ is the restore duration $D_R$ plus the delay until the first restore $\Delta r_0$. From  (\ref{firstdelay}), $\Delta r_0$ has mean  132 min and standard deviation 92.4 min.
  Then we can evaluate event duration mean and standard deviation from (\ref{meanD}) and (\ref{SDD}).
 For example, if there are $n=10$ outages, then 
 the event duration has mean 660 min and a standard deviation of 230 min.  
 If there are $n=100$ outages, then 
 the event duration has mean 3171 min and a standard deviation of 408 min.

When an outage event occurs, the primary question the customers want an answer for is ``How long will the power be out?".  
To help determine what should be announced to the public to answer this question, it is useful to estimate an upper bound on the restore duration that will be satisfied with a specified confidence level.

For each given value of number of outages $n$, the restore duration $D_R$ approximately follows a gamma distribution. We can estimate from (6) and (7) the mean and standard deviation of $D_R$ and then calculate the gamma distribution with that mean and standard deviation. 
This method of estimation is called the method of moments.
We now give more details.
The gamma distribution has  probability density function  
\begin{align}
    f(x)= \frac{\beta^{\alpha}}{\Gamma (\alpha)}x^{\alpha-1} e^{\beta x},\quad x\ge0,
\end{align}
where $\Gamma(\cdot)$ is the gamma function, $\alpha$ is the shape parameter and $\beta$ is the rate parameter. 
The gamma distribution mean is $\alpha/\beta$ and the standard deviation is $\sqrt\alpha/\beta$.
For each $n$, we can evaluate (6) and (7) and solve 
\begin{align}
    \overline{D_R}&=\frac{\alpha}{\beta}\\
      \sigma(D_R) &= \frac{\sqrt\alpha}{\beta}
\end{align}
for $\alpha$  and $\beta$.
These parameters specify a gamma distribution $f(D_R)$ for each value of $n$. 
Then we can easily evaluate the 95th percentile $D_{R95}$ of the gamma distribution that satisfies $\int_0^{D_{R95}}f(x)dx=0.95$. $D_{R95}$ estimates an upper bound on the restore duration that is satisfied by the actual restore duration with probability 0.95.

\begin{figure}[htb]
	\centering
	\includegraphics[width=\columnwidth]{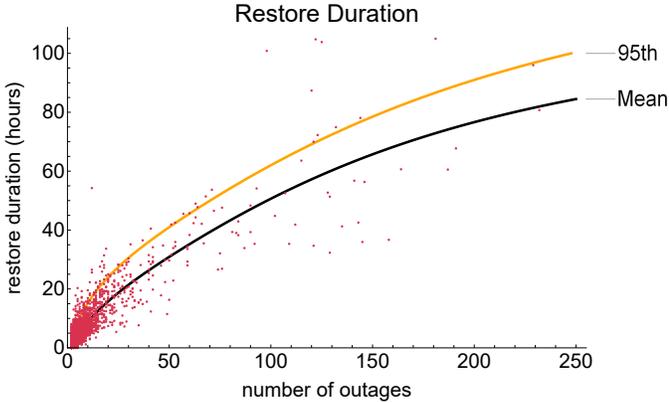}
	\caption{Curves show mean and 95th percentile of restore duration $D_R$   versus number of outages. Dots show the restore durations of  events in the data.}
	\label{restoremeanwith95th}
\end{figure}

The curves in 
Figure \ref{restoremeanwith95th} show an increasing and initially decelerating increase of mean restore duration as the number of outages increase, and a similar increase in the 95th percentile of restore duration. The dots in Figure~\ref{restoremeanwith95th} are the restore durations for the events in the data; they show how the mean and 95th percentile of restore duration calculated from the estimated gamma distribution summarize the empirical data. The event data become sparser as the number of outages increase.

\begin{figure}[htb]
	\centering
	\includegraphics[width=\columnwidth]{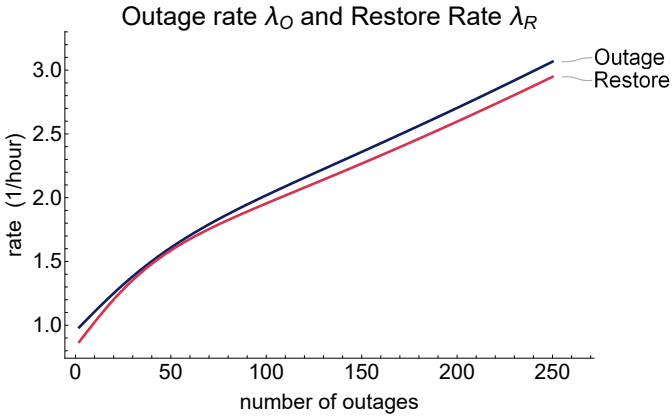}
	\caption{Outage rate $\lambda_O$ and restore rate $\lambda_R$ versus number of outages.}
	\label{ratecomparefig}
\end{figure}

Figure \ref{ratecomparefig} shows the outage rate $\lambda_O$ and the restore rate $\lambda_R$ obtained from (\ref{lambdaO}) and  (\ref{lambdaR}) as the number of outages varies. Both rates increase significantly as the number of outages increase. The outage rate results from the interaction of the weather with the grid, and depends on the design margin, age, and maintenance of the grid components. The restore rate results from the utility crews, restoration plans, and control room procedures. For up to 250 outages, the restore rate is quite close to the outage rate. The restore rate slightly lags the outage rate, showing the extent to which the utility restore process succeeds in keeping up with the outage process.

\begin{figure}[htb]
	\centering
	\includegraphics[width=\columnwidth]{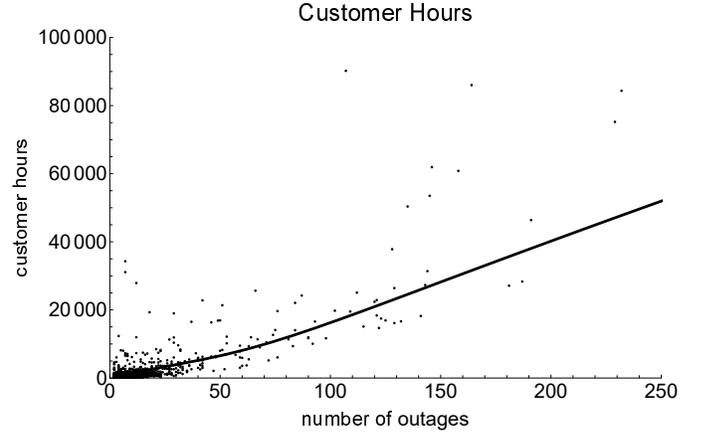}
	\caption{Curve shows mean customer hours $\overline{A}$  calculated from (\ref{Amean1}) versus number of outages. Dots show  customer hours $A$ of the events in the data.  }
	\label{customerareamean}
\end{figure}

The curve in
Figure \ref{customerareamean} shows the mean customer hours $\overline{A}$ calculated from (\ref{Amean1}) increasing as a function of the number of outages. 
The dots in Figure \ref{customerareamean} show the customer hours $A$ for each event in the data. There is considerable variability in the customer hours in the data for more than 100 outages. Future work could aim to analyze and quantify  this variability.

Although sections \ref{processes} and \ref{customerprocess} fit the data for the full range of our data up to 2000 outages, the data for the events with more than  250 outages become sparse and more variable.
In order to be cautious in our conclusions, we limit all the presented results in this section to events with up to  250 outages. 
Future work with more data or with more elaborate statistical methods might well extend the range of prediction.

\section{Discussion}
This section discusses the contributions of the paper in the larger context of power system resilience.

Our analysis is for distribution systems since the data are from a distribution system. 
However, the decomposition of resilience curves into outage and restore processes is general and  could be applied to resilience curves derived from data in transmission systems or in microgrids.
(Resilience curves for transmission systems are discussed in many papers, including \cite{NanRESS17,PanteliProcIEEE17,PanteliPS17,LuE18,OuyangSS12}, and resilience curves for microgrids are discussed in \cite{ZhouCST20, AndersonWilliamPhD20,MaPESGM19}.)
Note that the decomposition applies to resilience curves that track any quantity on the vertical axis; this paper is restricted to tracking numbers of components and  customers simply because these are the quantities available in our data.

Detailed outage data for transmission systems usually track number of components, and the corresponding outage and restore processes and associated metrics could be extracted. 
However, the details of the event definition and the specific forms of the processes could differ.
For example, we would expect the outage and restore processes for transmission systems to have some reduction in overlap, and their restore processes to often have an exponential recovery form.
There are also other approaches to quantifying resilience in transmission systems based on detailed outage data.
Resilience metrics are obtained by sampling transmission system outage data statistics  in \cite{KellyGorhamEPSR20}, and resilience is assessed by sampling from a Markovian influence graph driven by transmission system outage data in \cite{ZhouPMAPS20}.

There is extensive work with simulation, mitigation, optimization, and calculation of metrics based on detailed models of phases of resilience; for example, \cite{MaPS18,XuPS20,BieProcIEEE17,BhusalIEEEAccess20,WangPS16,OuyangSS12}. 
We see this detailed modeling as complementary to methods driven by historical data.
For example, to reduce customer hours lost, one could use formula (25) to identify how the   mean time between restores influences mean customer hours lost and then use a detailed simulation model of the restoration to show how mean time between restores could be reduced.

Our approach analyses resilience events with metrics at a systems level. The mechanisms of outage and restoration of individual components are complicated, and how individual outages and restorations and component design and maintenance and weather and restoration procedures interact is also complicated. 
The  outage and restore processes describe the aggregated outcomes of these complexities by tracking the number of components or  customers outaged or restored.
Note that which individual component is outaged or restored or the order of outage or restore is not tracked by the outage and restore processes.
We find that it is simpler and more direct to avoid description of individual component repair processes,
and this approach is made possible by the utility data that directly describe the outage and restore processes. 

Instead of analyzing individual resilience events, we analyze the statistics of many resilience events, as do \cite{CarringtonPESGM20,ChowPD96,ZapataTandDCE08,LiuRESS08,JaechPS19}.
It is certainly useful to analyze individual resilience events (e.g., \cite{ReedSYS09,WeiNNLS14}) and respond by proposing upgrades that would mitigate similar events. 
However, an exclusive focus on individual events may skew analysis and mitigation towards the last major event; quantifying resilience more broadly across many events is a useful complement.
Improving resilience is inherently a hard problem since the largest events with highest impact and significant risk are rare, and we conclude that analyses of both individual events and the statistics of all events is warranted.

Smaller distribution utilities may not have enough outages over a period of years to extract good statistics. 
    In that case, data could be combined from nearby distribution utilities with similar design practices and weather (for outage process metrics) or similar or shared restoration crews and procedures (for restore process metrics). 
    Also, government or reliability authorities could aggregate utility data to quantify resilience over a region to inform investments in resilience.

While the practical advantages such as realism and lack of modeling assumptions of analyzing historical data are clear, the limitations should also be noted.
The historical data are what did happen and a pragmatic sample of what could happen, but cannot have many samples of all the possible rare events. 
Moreover, some of the extremes of weather driving  resilience events will become gradually more frequent and more severe as the planet warms \cite{IPCC14}, so that the severity of some types of resilience events will increase.
This underlines the importance of developing multiple approaches to quantifying and improving resilience, including the approach developed in this paper.
Extracting resilience metrics from utility data and thus quantifying aspects of resilience can help to justify resilience improvements to mitigate disasters.

\section{Conclusions}

We process 5 years of distribution system outage data to extract and study many resilience events in which outages accumulate and are restored. 
As appropriate for quantifying resilience, we focus only on the resilience events, and do not analyze the frequency of these events or the times between events that are of interest in other kinds of reliability analysis.

It is usual to separate resilience curves for events into successive non-overlapping phases in time such as outage and recovery.
However, our distribution utility data show that   outage and restore processes  typically occur together for most of the event.
Therefore,  instead of using successive phases,  we show how resilience curves tracking the number of outaged components or customers  can be easily decomposed into outage and restore processes that can occur at the same time.
 These outage and restore processes describe the same information as the resilience curve, but usefully correspond to different aspects of resilience: the outages are caused by weather interacting with the strength of the components, whereas the 
restores are done by utility crews and control rooms. 
The decomposition of the resilience curve into outage and restore processes is known as the Jordan decomposition in mathematics.

We compute some basic statistics of the outage and restore processes. 
In particular, we estimate fits for the mean and standard deviations of the times between outages and the times between restores as functions of the number of the outages. The function fitting has the effect of smoothing and interpolating the noisy data.  We also  estimate the means and standard deviations of the number of customers outaged and of the delay until the restore process starts.

Then, given the predicted number of outages, which is estimated in several ways  by previous work \cite{LiuPS07,LiuJIS05,KankanalaPS14,ZhuEPSR07,ZhouPS06,AlvehagPD11,LiIBM10}, we  obtain formulas for the means of  standard resilience metrics, such as restore and event  durations, restore and outage rates, and the customer hours lost. 
These formulas for standard resilience metrics usefully quantify the resilience processes and show how the metrics depend on the number of outages,  the delay before restoration starts, the average time between restores, and the average number of customers disconnected per outage.
The outage rate quantifies the overall grid fragility under weather stress and the restore rate quantifies the overall  performance of utility crews. 
This quantification can help inform investments that improve these metrics. 
We also estimate the standard deviations of the restore duration and the event duration. 
This leads to estimates of probable upper bounds of restore durations. 
These credible upper bounds based on past performance should be useful to utilities for informing customers about their outage duration with confidence.
The utility data available to us limited our resilience processes to counting outages of components or customers. If available,  quantities such as power outaged could be  similarly analyzed to obtain useful metrics such as energy not served.

Our approach models the outage and restore processes directly from utility data and avoids the complexities of modeling individual component restoration or repair times and their order of restoration.
That is, since the utility data itself incorporate the detailed complexities of resilience, we can give a high-level description  and quantification of  resilience.
This high-level data-driven approach is very much a useful  complement to the detailed modeling of the restoration processes by other authors.

In summary, we extract and separate the outage and restore processes from distribution utility outage data in a new way, estimate 
the statistics of times between successive restores or outages, and then show how standard resilience metrics can 
be derived from these statistics. 
The overall effect is to 
compute some useful resilience metrics from practical utility data.

   \vskip -2.2\baselineskip plus -1fil

 \begin{IEEEbiography}[{\includegraphics[width=1in,height=1.25in,clip,keepaspectratio]{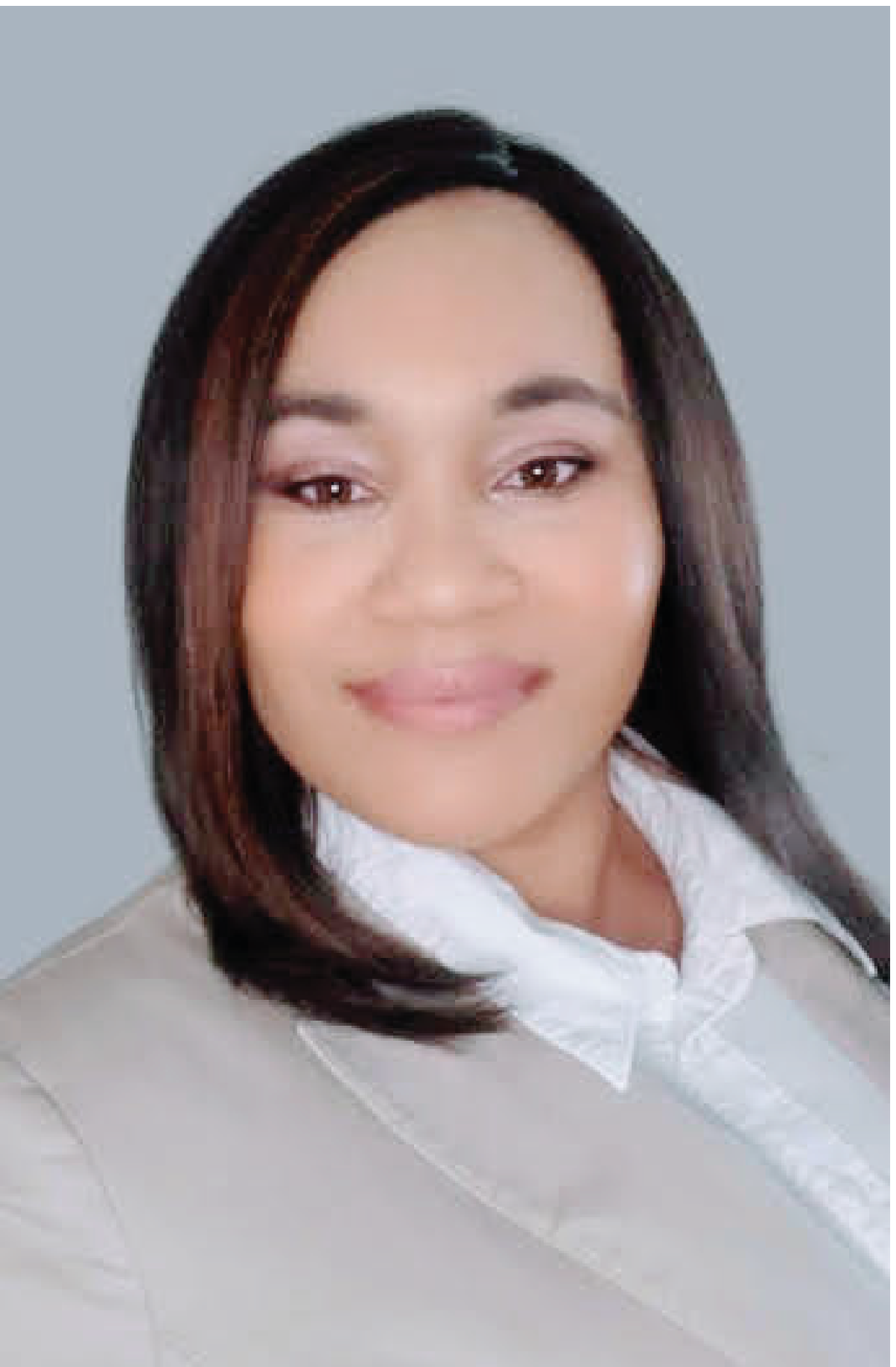}}]{Nichelle'Le K. Carrington}
    (S‘18)  received the B.S. degree in electrical engineering in 2014 from North Carolina Agricultural and Technical State University, Greensboro NC, USA. She is currently pursuing the Ph.D. in electrical engineering with a focus in power and energy systems at Iowa State University.  Her current research interests include distribution system resilience, analysis and assessment of real utility data to extract resilience metrics, data analytics in smart grids, and industry applications leveraging power system smart meter data.
    \end{IEEEbiography}
    
      \vskip -2.2\baselineskip plus -1fil
    
   \begin{IEEEbiographynophoto}{Ian Dobson}
   (F’06) received the B.A. degree in mathematics from Cambridge University and the Ph.D. degree in electrical engineering from Cornell University.
He previously worked for British industry and  the University of Wisconsin-Madison, and is currently Sandbulte Professor of Electrical  Engineering at Iowa State University. His interests are blackout risk,  power system stability, synchrophasors, complex systems, and nonlinear dynamics.
    \end{IEEEbiographynophoto}
    
        \vskip -2.2\baselineskip plus -1fil
    
 \begin{IEEEbiographynophoto}{Zhaoyu Wang}
 (S’13-M’15-SM’20) is the Harpole-Pentair Assistant Professor with Iowa State University. He received the B.S. degree from Shanghai Jiaotong University, and Ph.D. degree from Georgia Institute of Technology. Dr. Wang has been a recipient of the National Science Foundation (NSF) CAREER Award and the IEEE PES Outstanding Young Engineer Award. 
 His research interests include optimization and data analytics in power distribution systems and microgrids. 
    \end{IEEEbiographynophoto}

\end{document}